\documentclass[12pt]{article}
\def\half{{1 \over 2}}
\usepackage{epsfig,latexsym}

\begin{document}
\hfill Bicocca-FT-01-06 \vskip .1in \hfill hep-th/0101232

\hfill

\vspace{20pt}

\begin{center}
{\Large \textbf{ RG FLOWS TOWARDS IR ISOLATED FIXED POINTS:}}
{\Large \textbf{SOME TYPE 0 SAMPLES }}
\end{center}

\vspace{6pt}

\begin{center}
\textsl{F. Bigazzi} \footnote{e-mail: francesco.bigazzi@mi.infn.it} \vspace{20pt}

\textit{Universit\`{a} di Milano-Bicocca, Dipartimento di Fisica}

\end{center}
                                     
\vspace{12pt}
\begin{center}
\textbf{Abstract }
\end{center}

\vspace{4pt} {\small \noindent
We perform here a critical analysis of some non-supersymmetric gravity
 solutions in Type 0B string theory.
We first consider the most general configuration of parallel $N$ electric  
and $M$ magnetic $D3$-branes. The field theory living on their worldvolume is
non-supersymmetric and non-conformal (if $N \ne M$) and has gauge group
 $SU(N)\times SU(M)$. We study the IR regime of the conjectured dual gravity
 background. A fine tuned solution exists with an asymptotically vanishing
 tachyon and a running dilaton, which could correspond to a flow towards
an IR isolated fixed point at strong coupling.
 This opens the question upon the possibility of extending AdS/CFT techniques
 to flows towards  IR isolated  fixed points.
We then use D3-branes as probes of this and other Type 0 backgrounds
available in literature which cover the $M=0$ (or $N=0$) case. We shaw that for the fine tuned IR conformal solutions the stacks of parallel branes are plagued by instabilities, due to the repulsive force between branes of the same type. Curiously for a particular solution which could be dual to a confining gauge theory, the stack should instead be stable.
\vfill\eject
\noindent
}
\section{Introduction}
Attempts to extend the $AdS/CFT$ correspondence to non-supersymmetric and/or
non-conformal theories have been presented in the last years. Possible string duals of nonconformal theories can be obtained in a supersymmetric context by considering deformations of conformal backgrounds
 \cite{gppz,freed} or using fractional branes \cite{kn,ktfrac,ks,mn}. These backgrounds are dual to strongly coupled gauge theories. The IR dynamics of such theories can be different from the familiar one of their weakly coupled cousins, 
but they are nevertheless expected to exhibit familiar phenomena as 
confinement, chiral symmetry breaking etc. The IR region of the 
supergravity background, which should describe these phenomena, is
generically plagued with naked singularities, but regular solutions
are known nowadays \cite{ks}.
 
Type 0 theory could provide the natural arena to extend these constructions to
non-supersymmetric theories. In \cite{bgz}, for example, a Type 0 regular 
solution was found which is a possible non-supersymmetric version of the one 
presented in \cite{ks}.

Type 0 closed strings have a doubled set of RR fields compared to their type II cousins, 
hence they also possess twice as many D-branes \cite{kt}. For example,  type 0B
contains a RR  4-form gauge potential whose field strength is not restricted to be
selfdual. This implies that there are two types of D3-branes: 
those that couple electrically to this gauge potential, and those that couple 
magnetically.

In the closed string spectrum there is also a tachyon of 
$m^2=-{2\over\alpha'}$ which is not removed by the non-chiral GSO 
projection $(-1)^{F+\tilde F}=1$ responsible for the absence of spacetime supersymmetry in the theory.

Despite the presence of a tachyon in the bulk we can 
construct stable D-branes in Type 0 (see for example \cite{berg} for a boundary state
construction and \cite{costa} for a study of intersecting Type 0 branes). Here ``stable'' means that open strings starting
and ending on these branes do not contain tachyons. This means that no evident
instability appears for the  gauge theories living on the D-brane worldvolume. In \cite{kt1,k} it was argued that this may imply a nonperturbative stabilization from the string theory side. 

The first attempts to construct Type 0 duals of large $N$ nonsupersymmetric
4-dimensional gauge theories were presented in \cite{kt}. There, a stack of 
$N>>1$ parallel electric (or magnetic) branes is considered. The worldvolume 
gauge theory is $SU(N)$ coupled with 6 adjoint scalars. The tachyon may be stabilized by RR flux. The conjectured dual gravity background asymptotes to 
$AdS_5\times \hat S^5$ in the UV. In this region the tachyon is supposed to relax towards a constant value and the dilaton runs logarithmically. This corresponds to the expected asymptotic freedom of the gauge theory. In the IR we can have a fine tuned conformal solution \cite{kt0} or a more general one \cite{minahan} which could correspond to a confining gauge theory.
 
An important question upon such a construction, is how can we construct the
background. In fact we can consider the branes stacked parallel to each other 
as a stable configuration if, for example, no repulsive force is experienced
by them. In \cite{z}, \cite{tz} the force between electric branes was 
extrapolated from the gauge theory scalar potential at 1-loop. Clearly this kind of analysis is valid only in the UV region (long distances from the branes).

In \cite{kt1} it was considered a stack of $N$ electric and $N$ magnetic D3-branes . For such a stack the net tachyon tadpole cancels so that there
exists a classical solution with $T=0$. In fact, since the stack couples
to the selfdual part of the 5-form field strength, the type 0B 3-brane classical solution is identical to the type IIB one. This may suggests that 
the low-energy field theory on $N$ electric and $N$ magnetic D3-branes is dual to the $AdS_5\times S^5$ background of type 0B theory and is therefore conformal in the planar limit. This theory is the $SU(N)\times SU(N)$ gauge theory coupled to 6 adjoint scalars of the first $SU(N)$, 6 adjoint scalars of the second $SU(N)$, and fermions in the bifundamental representations (4 Weyl fermions in the $(N, \bar N)$ and 4 Weyl fermions in the $(\bar N, N)$). This theory is a regular $Z_2$ orbifold projection of the ${\cal N}=4$ $SU(2N)$ gauge 
theory \cite{kt1}, \cite{ns}. The conjectured dual background is stable for small or intermediate
values of the `t Hooft coupling, which corresponds to large curvature from the gravity 
side. The AdS/CFT correspondence applied to this particular case implies \cite{k} that, 
as $\lambda = g_{YM}^2 N$ is increased, a phase transition should happen at a critical 
value $\lambda_c$ rendering the theory unstable.
In \cite{tz} the stability of this ``selfdual'' stack was considered, again
extracting informations from the UV gauge theory.

In this paper we try to extend these constructions to the  general case in which
N electric and M magnetic branes are stacked parallel to each other. The field theory living on their worldvolume has gauge group $SU(N)\times SU(M)$ with adjoint scalars and 
bifundamental fermions in the $(\bar N,M)$, and in the $(N, \bar M)$.
This theory is not expected to have asymptotic freedom.

In sections 2 and 3  we examine , from the AdS/CFT point of view, the IR of the theory, showing that a fine tuned gravity solution exists, corresponding to an IR isolated conformal fixed point.
 
This opens the question upon the possibility of extending AdS/CFT 
techniques to these particular cases. In the selfdual case in which $N=M$ 
we have a line of fixed points. We can apply the standard techniques to extract,
for example, the conformal dimensions of the gauge theory operators 
corresponding to the gravity fields. In the general case we find that
a certain regular behaviour seems to be present (despite the fact that the
 dilaton blows up). This could be helpful in trying to calculate correlator
functions and operator dimensions also in the case of the isolated CFT. For related
issues on RG flows in Type 0 contexts, see \cite{angel}.

In section 4 we examine the stability of our gravity solution and of the ones presented in \cite{kt0},\cite{minahan} which cover the $M=0$ case.  
D-branes as probes will serve this purpose. In this way we will try to compute
the force felt by a brane due to the interaction with the others in a string context, so extending the results in \cite{tz}. We will show that, as expected,
 Type 0 D-brane stacks are in general IR instable configurations, due to the repulsive nature of the forces between branes of the same type. This fact may suggest that configurations of branes more complicate than the simple stack of parallel ones could take part in the game. We will also show that for a particular confining solution found in \cite{minahan} the stack could instead be stable.
Clearly, the absence of supersymmetry and the fact that string loop corrections 
could become important in the IR regime, render this kind of analysis
speculative.

Using D-branes as probes will also furnish us the natural arena for discussing some issues related to holography and radius/energy relation in nonconformal contexts.
\section{Gravity duals of non-susy $SU(N)\times SU(M)$ gauge theories}
Let us consider the classical background produced by a stack of parallel
$N$ electric and $M$ magnetic $D3$-branes in Type $0B$. The world-volume supports the gauge theory $SU(N)\times SU(M)$ with bifundamental fermions in the
$(\bar N,M), (N, \bar M)$ and adjoint scalars.

We use the conventions in \cite{kt} for the string frame metric ansatz: 
\begin{equation}
ds^2~=~e^{\half\phi}\left(e^{\half\xi-5\eta}d\rho^2~+~e^{-\half\xi}dx_{||}^2
~+~e^{\half\xi-\eta}d\Omega_5^2\right),
\label{met}
\end{equation}
where $\phi$, $\xi$ and $\eta$ are functions of $\rho$ only 
(the same applies to  $T$ and $C_4$). From this
 we can rewrite the ten dimensional equations of motion
for the metric, dilaton, RR fields, and tachyon so that they can be deduced
from the subsequent effective action (the dots stands for derivatives with respect to $\rho$):
\begin{eqnarray}
S&=&\int d\rho\left[\half \dot\phi^2+\half\dot\xi^2+{1\over4}\dot T^2
-5\dot\eta^2- V(\phi,\xi,\eta,T)\right],\\
V(\phi,\xi,\eta,T)&=&g(T)e^{\half\phi+\half\xi-5\eta}+20e^{-4\eta}-
h(T)e^{-2\xi},
\label{toda}
\end{eqnarray}
\noindent and the constraint:
\begin{equation}
\half \dot\phi^2+\half\dot\xi^2+{1\over4}\dot T^2-5\dot\eta^2+
V(\phi,\xi,\eta,T)~=~0.
\label{constr}
\end{equation}
\noindent Here:
\begin{equation}
g(T)~=~ -{1\over4}{m^2}T^2-{\lambda}T^4+O(T^6);\quad   ({m^2= {-2\over{\alpha'}})}, 
\end{equation}
\noindent is the tachyon potential, and
\begin{equation} 
h(T)~=~Q^2{f(T)}^{-1}+P^2f(T) 
\end{equation} 
\noindent is the RR-tachyon coupling( $Q\approx N$ and $P\approx M$ are the electric and magnetic charges, respectively).

In \cite{kt} was found that, near $T=0$, 
\begin{equation} 
f(T)~=~1+T+\half T^2+O(T^3). 
\label{effe}
\end{equation} 
\noindent Recently it was argued in \cite{t} that in the low energy effective 
action of Type 0 string theory no non-derivative contributions to the tachyon potential appear, leaving us with only the mass term ($\alpha'=1$): 

\begin{equation} 
g(T)~=~ \half T^2. 
\end{equation} 

\noindent Now we attempt to study the possible ( asymptotic) solutions of the equations of motion deduced from the action (\ref{toda}). We find a gravity solution which could correspond to
 an IR fixed point from the gauge theory side.

It is understood that in the UV, where the gravity approximation is no longer 
valid ($\alpha'$ corrections are not negligible in general), results must be 
taken {\it cum grano salis}.


\subsection{IR solution}

The equations of motion derived from the action (\ref{toda}) are:

\begin{eqnarray}
\ddot\xi~+~\half g(T)e^{\half\phi+\half\xi-5\eta}~+~2h(T)e^{-2\xi}
~=~0 , \nonumber \\
\ddot\eta~+~\half g(T)e^{\half\phi+\half\xi-5\eta}~+~8e^{-4\eta}~=~0, \nonumber \\
\ddot\phi~+~\half g(T)e^{\half\phi+\half\xi-5\eta}~=~0, \nonumber \\
\ddot T~+~2 g'(T)e^{\half\phi+\half\xi-5\eta}~-~2h'(T)e^{-2\xi}~=~0.
\label{eqmot}
\end{eqnarray}

\noindent Now let us look for IR asymptotic solutions of these equations which are fine tuned so to have $T\approx 0$ as asymptotic solution.
Note that $T=0$  cannot be an exact solution like in the dyonic case \cite{kt1}.
This is because we are considering the most general case in which $Q\ne P$.
In fact we have:
\begin{equation}
h'(T)~=~\left[{-{Q^2\over{f(T)^2}}}+P^2\right] f'(T)
\end{equation}
and, because near $T\approx 0$ we have $f'(T)\approx 1$, we find
\begin{equation}
h'(T\approx 0)\approx(P^2 - Q^2).
\end{equation}

So we find that we can have $T=0$ as an exact solution of the equations of motion only if $Q=P$ which is the case considered in \cite{kt1}. It corresponds to an $AdS_5\times S^5$
background which is conjectured to be dual to a large N non-supersymmetric conformal 
 field theory with gauge group $SU(N)\times SU(N)$.
  
In the most general case $Q\ne P$, we can have a zero tachyon as asymptotic solution  
in the regime in which:
\begin{equation}
e^{-2\xi}\rightarrow 0,
\end{equation}

\noindent which corresponds to taking $\rho\rightarrow\infty$ (IR limit from the gauge
 theory point of view).

In fact a consistent set of asymptotic solutions of (\ref{eqmot}) is:
\begin{eqnarray}
\xi\approx \log \sqrt{2(Q^2 +P^2)} + \log\rho , \nonumber \\
\eta\approx \log 2 + \half \log\rho , \nonumber \\
\phi\approx  -\log \sqrt{2(Q^2 +P^2)} + 2\log(\log\rho), \nonumber \\
T \approx {-{16\over{\log\rho}}{(Q^2-P^2)\over{(Q^2+P^2)}}}.
\label{eqgen} 
\end{eqnarray}

\noindent They reduce to the first terms of the solutions found in \cite{kt}, \cite{kt0}
in the case $P=0$. When $P=Q$ we find a zero tachyon which is also an exact
 solution \cite{kt1}. Just like in \cite{kt0} we obtain here ,in the IR, 
 a solution which becomes $AdS_5\times S^5$ and in which the dilaton blows 
 up (string loop corrections are not negligible). We could interpret this, as in
\cite{kt0}, as an isolated CFT at infinite coupling.

The situation in the UV is much less under control in the sense that we need 
to know the details of the coupling function $f(T)$ and of the tachyon potential $g(T)$.

\section{Asymptotically $AdS$ backgrounds and RG flows}
In this  section we comment about the possibility of extending the techniques developed in \cite{gppz} and
\cite{freed}, to RG flows towards IR isolated fixed points. 
The gravity solution we consider here is the general Type 0 one previously 
discussed.

First of all we perform a change of variables in order to write the 10-dimensional metric (\ref{met}) in the Einstein frame in the following form (we are using here the same notations as in \cite{ktfrac}):

\begin{equation}
ds^2~=~e^{-5q(\tau)}\left[{d\tau}^2 + e^{2A(\tau)}dx^{\mu}dx^{\nu}\eta_{\mu\nu}\right] +e^{3q(\tau)}{d\Omega_5}^2 .
\end{equation}

This is realized by defining:
\begin{eqnarray}
A~\equiv~{\xi\over6}-{5\over6}\eta ,\nonumber \\
q~\equiv~{\xi\over6}-{\eta\over3}, \nonumber \\
{d\tau\over{d\rho}}~=~e^{4A}.
\end{eqnarray}

In this way we can rewrite the radial effective action (\ref{toda}) as:
\begin{equation}
S_5~=~-4 \int d\tau e^{4A}\left[3{\dot A}^2 -\half G_{ab}\partial\phi^a\partial\phi^b - V(\phi)\right],
\label{5da}
\end{equation}
where:
\begin{equation}
G_{ab}\partial\phi^a\partial\phi^b~=~15{\dot q}^2 +{{\dot\phi}^2\over4}+{{\dot T}^2\over 8},
\end{equation}
\begin{equation}
V(\phi)~=~-5e^{-8q}+{h(T)\over 4}e^{-20q}- {T^2\over 8}e^{\half \phi -5q},
\end{equation}
and
\begin{equation}
\phi^a~=~(q,\phi,T).
\end{equation}
As a difference with \cite{ktfrac} the theory is now non-supersymmetric and we are not 
able to find any superpotential. By shifting $q$, $\tau$ and $\phi$ as
\begin{eqnarray}
q\rightarrow {\sqrt{15}}q - {{\sqrt{15}}\over4}\left({1\over3}\log(Q^2+P^2) -\log2\right), \nonumber \\
\phi\rightarrow \half\phi + {1\over4}\left(\log(Q^2+P^2) -3\log2\right), \nonumber \\
\tau\rightarrow 2(Q^2+P^2)^{-{1\over3}}\tau ,
\end{eqnarray}
we can put the potential in the form (we use the same symbols for the 
new variables):
\begin{equation}
V~=~ -5e^{-8q} + 2e^{-20q}\left( 1+ T {{P^2-Q^2}\over P^2 + Q^2} + \half T^2 + ...\right)
- {T^2\over 8}e^{\half \phi -5q}.
\end{equation}
Consider first the dyonic case of \cite{kt1}; for $P=Q$, $T=0$, $\phi=const$ and we have
the $SU(N)\times SU(N)$ CFT with a line of fixed points.
Expanding the action (\ref{5da}) near $q=0$, which is a critical point for the potential,  we can deduce for $q$ the effective lagrangian:
\begin{equation}
L_{q}~\approx~ -\half{\dot q}^2 +3 -{\half 32}{q}^2.
\label{fl}
\end{equation}
In the normalizations where $V_{min}=-3$, the relation mass/conformal dimension is:
\begin{equation}
m^2~=~\Delta(\Delta - 4).
\label{delta}
\end{equation}
From (\ref{fl}) we get $m_{q}^2=32$: thus $q$ corresponds via the $AdS/CFT$ map (\ref{delta}) to the source for a 4d operator of dimension $\Delta=8$, which we 
identify with $tr(F^2\bar F^2)$. Analogous results were found in the conifold case of \cite{bgz}.
As usual \cite{k} $\phi$ is the source for ${F_1}^2 + {F_2}^2$ while $T$ 
is the one for ${F_1}^2 - {F_2}^2$. For the mass squared of $T$ we get:
\begin{equation}
m^2_T~=~ (16 -2{\sqrt x}),\quad  e^\phi= x
\end{equation}
which is the result in \cite{k}, showing that the tachyon is unstable for large
'tHooft couplings.

For $P\neq Q$ we have an asymptotic solution corresponding to (\ref{eqgen}): 
$\phi \approx 2 \log |\tau|$, $T\approx \tau^{-1}$ for $\tau\rightarrow -\infty$. 
Notice that the dilaton diverges, but $ T^2 exp(\half\phi -5q)\rightarrow 0$, showing that the potential is finite in the limit. Here $q=0$ is still an $AdS_5$ minimum, suggesting the existence of a CFT in the IR at strong coupling.

Little is known about $AdS/CFT$ in the case of isolated fixed points in four dimensions
at strong coupling. It would be interesting to understand better the rules of $AdS/CFT$
in the case here examined. In particular it would be interesting to compute the correlation functions and the spectrum. $V(q)$ is regular in the limit $T\rightarrow 0$, 
$\phi\rightarrow\infty$, so we could extract ${m_q}^2 = 32$. It would be interesting to compute the dimension of the operators corresponding to $T, \phi$ whose fields diverge.



\section{D3-branes as probes of Type 0 backgrounds}

Previously we considered gravity solutions taking the hypothesis that they
are generated by a stack of parallel branes. This assumption has a sense only 
if we can construct such a configuration, or, in other words, if that configuration is stable. In \cite{tz} the interaction force between electric
(or magnetic) branes was estimated by using the gauge theory approximation at one
loop. So those results should be taken seriously only in the UV region,
 in which the gauge theory is weakly coupled. Here we reproduce
those results and try to extend them to the IR region following another strategy.

We perform, in fact, an analysis of the backgrounds in \cite{kt0}, \cite{minahan} which cover the $M=0$ case, and of the general one previously discussed, using D-branes as probes.

We start by reviewing  some general technical details and examine the relations between the UV probe calculations and the ones examined in \cite{tz}. Then we examine the IR region of the backgrounds cited above and show that the stacks of parallel branes are in general plagued by instabilities due to the repulsive nature of the force felt by the probe. This suggests that the branes could be assembled in more complicate configurations (such as spherical shells)
than the simplest one in which they are stacked parallel to each other.
We will also suggest that for a particular confining solution found in \cite{minahan} the stack of parallel branes could be instead a stable configuration.

For a review of D-branes as probes of (super)gravity backgrounds see \cite{j}.

\subsection{General considerations}
 The DBI and CS parts of the action for a D3-electric brane probe (in a background in which no 3-form RR or NSNS field strength is turned on) are \cite{garousi}:
\begin{equation}
{S_p}~=~-{\tau_p}\int {{d\xi}^4}{e^{-\phi}}K(T)\sqrt{-\hat{G}}~+~{{\mu}_p}\int {\hat C}_4 ,
\label{dbi}
\end{equation}

\noindent where:
\begin{equation}
{\hat G}_{\alpha\beta}= {{\partial x^M}
\over{\partial \xi ^\alpha}}{{\partial x^N}\over{ \partial \xi ^\beta}}g_{MN},
\end{equation}
\noindent is the induced metric on the D-brane ($\alpha, \beta = 0, 1, 2, 3$ are the worldvolume coordinates while $M, N = 0,...,9$ are the spacetime ones). Here $g_{MN}$ is the string frame metric (\ref{met}).
The function $K(T)$ which couples the probe to the bulk tachyon has the following
expansion near $T=0$ \cite{garousi}:
\begin{equation}
K(T)~=~1+ {T\over4} + {3\over{32}}T^2 + O(T^3).
\label{k(t)}
\end{equation}
\noindent We now use the static gauge: $x^0=t; {x^{1,2,3}}={\xi^{1,2,3}}; {x^m}={x^m}(t), m=4,...9$, so that we can rewrite (\ref{dbi}) as:
\begin{equation}
{S_p}~=~-{\tau_p}{V_p}\int dt K(T)g^2_{00} \sqrt{ 1+ {g_{mn}\over{g_{00}}}{\dot x}^m{\dot x}^n }~+~ {{\mu}_p}{V_p} \int dt {\hat C}_4,
\label{probact}
\end{equation}

\noindent Now we can rewrite the metric (\ref{met}) in the Einstein frame, in the more familiar form:

\begin{equation}
ds^2~=~ e^{2A(u)}dx^{\mu}dx^{\nu}\eta_{\mu\nu} + e^{2B(u)}\left( du^2 +u^2d{\Omega_5}^2 \right),
\label{umetr}
\end{equation}

\noindent where 
\begin{equation}
{\partial\rho\over\partial{log u}}~=~- e^{2\eta},
\end{equation} 

\noindent and
\begin{eqnarray}
A \equiv -{\xi\over 4}, \nonumber \\
B \equiv {\xi\over 4} - {\eta\over 2} - \log u.
\end{eqnarray}

\noindent From (\ref{probact}) we can thus extract the effective lagrangian:
\begin{equation}
L~=~ -mc^2\sqrt{ 1 - {v^2\over{c^2}}} - V,
\end{equation}

\noindent where $v$ is the transverse velocity and
\begin{eqnarray}
c~\equiv~ e^{A-B}, \nonumber \\
m~\equiv~ {\tau_p} K(T)e^{2(A+B)}, \nonumber \\
V~\equiv~ -\mu_p{\hat C}_4 ,
\label{c}
\end{eqnarray}
are all functions of $u$.

As usual in D-brane probe analysis we will take the low velocity limit ${v\over c} <<1$.
As pointed out in \cite{minic} this requires some care being the function $c$
 not constant (this is true also in the supersymmetric BPS case). With this in mind,
we find the low velocity lagrangian:
\begin{equation}
L~=~ E_k - V,
\label{lagpr}
\end{equation}
where the effective kinetic energy is:
\begin{equation}
E_k~=~\half {\tau_p}K(T){e^{-\eta}\over u^2} v^2 .
\label{cinet}
\end{equation}
To write explicitly the potential, let us consider for example the $M=0$
case. In the ansatz of \cite{kt}, we have, for the electric RR 5-form 
field strength:
\begin{eqnarray}
{F_5}(\rho) = d[ A(\rho)d{x^0}\wedge d{x^1}\wedge d{x^2} \wedge d{x^3} ]\\
{\dot A}(\rho)= -{g_s}^{-1}2Q{e^{-2\xi}}f^{-1}(T).
\end{eqnarray}
By using this result, we see that the potential term in (\ref{lagpr}) takes the form 
($\tau_p = {g_s}^{-1}\mu_p$):
\begin{equation}
V~=~ \tau_p\left[ K(T)e^{-\xi} + 2Q\int d\rho e^{-2\xi}f^{-1}(T)\right],
\label{pot}
\end{equation}
which we shall rewrite as a function of $u$.

Let us end this introductory section with a series of observations. In static gauge,
the probe effective action may be written as (let us take $\dot\theta_i=0$):
\begin{equation}
S\approx \tau_pV_p\int dt \left[K(T)e^{-\phi}\left[ {e^{\phi - \eta}\over u^2}{(\partial u)}^2 + F^2\right] - V(u)\right],
\end{equation}
where we have inserted explicitly the U(1) field strength living on the brane.
From this we could extract the formal gauge theory lagrangian:
\begin{equation}
L = {1\over{g^2}_{YM}}\left[ (\partial\Phi)^2 + F^2 - V_f(\Phi)\right]
\end{equation}
after having defined:
\begin{equation}
d\Phi= {e^{\half(\phi - \eta)}\over u} du
\label{PHI}
\end{equation}
with the identifications:
\begin{eqnarray}
{1\over {g^2}_{YM}}= K(T)e^{-\phi}, \nonumber \\
{1\over {g^2}_{YM}}V_f = V .
\label{naive}
\end{eqnarray}   
\subsubsection{Matching with gauge theory results in the UV}
In order to review the physical relevance of the considerations above,
let us briefly recall the details of the UV ( $\rho\rightarrow 0$) solution corresponding to a stack of N parallel electric D3-branes \cite{kt0,minahan}. Here we take
$g(T)=\half T^2$ and suppose f(T) has a minimum for $T=T_{UV}$.
\footnote{ For example this could be obtained with:
\begin{equation}
f(T)=e^{T-{T^3\over 3}}
\end{equation}
which has a minimun for $T=-1$ and reproduce the near $T=0$ behaviour (\ref{effe}).}
So, with $\rho=e^{-y}$, we have \cite{minahan,maggiore}:

\begin{eqnarray}
T=T_{UV}+{8\over T_{UV}y}+ {4\over y^2}\left(39\log y + k -20\right) + O\left({\log^2y\over y^3}\right),
\nonumber \\
\xi={\log \sqrt{2f^{-1}(T_{UV})}Q} -y + {1\over y}+ {1\over 2y^2}\left(39{\log y} + k -104\right) + O\left({\log^2y\over y^3}\right),\nonumber \\
\eta= -{\half y}+\log 2 + {1\over y} + {1\over2y^2}\left( 39\log y  + k - 38\right) + O\left({\log^2y\over y^3}\right), \nonumber \\
\phi
 = -2\log(C_{0}Q) -2\log y +15\log2 +{1\over y}\left(39\log y + k\right) +O\left({\log^2y\over y^3}\right).
\label{uvsol}
\end{eqnarray}
In the near UV we thus have:
\begin{equation}
\log\rho + \log(\log\rho)^2 \approx -4\log u, \quad  u\rightarrow\infty \quad  (\rho \rightarrow 0) .
\end{equation}
The kinetic term thus asymptotes to:
\begin{equation}
dE_k~=~{\half \tau_p} K(T)4\log u \left(du^2 + u^2 d{\Omega^2}_5 \right).
\end{equation}
The behaviour of this term depends also from that of $K(T)$. If $K(T)$ is asymptotically constant in the UV we get a divergent kinetic term. But what is the real behaviour of $K(T)$? At this stage of approximations nothing sensible can be said about that. Nevertheless, if we argument, as in \cite{kt}, that gravity approximation could capture some gauge theory physics, we can tentatively extract some informations from the gauge theory side. For example we know that 
\cite{kt1} :
\begin{equation}
g^{-2}_{YM}\approx log u .
\end{equation}
By using the relation in (\ref{naive}) and observing that $e^{-\phi}\approx (\log u)^2$ we can tentatively predict $K(T)\approx (\log u)^{-1}$.
\footnote{If, for example  $K(T)=sinh[a(T-T_{UV})]$, we could have this asymptotic behaviour in the UV and a near $T=0$ behaviour similar to that in (\ref{k(t)}).}
From this we could have a constant kinetic term for the probe.

For the potential term (\ref{pot}) in the probe effective action we have the following near UV behaviour:
\begin{equation}
V~\approx~ {\tau_{p}\over Q}\left[{{K(T)-\sqrt{2f^{-1}(T_{UV})}}\over\sqrt{2f^{-1}(T_{UV})}}\right] u^4\log^2u,
\end{equation}
which tends to $\pm\infty$ for $u\rightarrow\infty$, depending on the details of $K(T)$ and $f(T)$. From these details, thus depends the nature of the force that the probe experts (if attractive or repulsive).
By identifying $V$ with the scalar potential term in the dual field theory as in (\ref{naive}) we can predict the following functional form for $V_f(\Phi)$:
\begin{equation}
V_f(\Phi)~\approx~ \Phi^4\log{\Phi\over\Lambda},
\end{equation}
a result which  was obtained from 1-loop calculations in field theory \cite{z}.

Let us finally observe that the slow velocity limit  for the probe action, is a sensible
one in the UV because, from (\ref{uvsol}) and (\ref{c}), we can deduce that $c\rightarrow\infty$ when $u\rightarrow\infty$. This could be responsible for the correct behaviour of the potential term for scalars fields, despite the gravity approximation is not valid in the UV.

\subsection{Probing the $M=0$ IR solutions}

 Let us now analyse the IR solutions discussed in \cite{kt0} and \cite{minahan} for the case $M=0$.
In the first case we have a fixed point at infinite coupling , while in the latter we have a confining solution.
In the conformal case \cite{kt0} the IR behaviours are ($\rho=e^y \rightarrow \infty$):
\begin{eqnarray}
T~=~ - {16\over y}
 -  {  8\over y^2}   ( 9 \log y  - 3) + O ( { \log^2 y \over y^3}), \nonumber\\  
\xi~ =~  \half \log (2 Q^2)  +  y   +    {9\over y}
 +   {  9\over 2 y^2}   ( 9 \log y  - {20\over 9}
) + O ( { \log^2 y \over y^3}), \nonumber\\
\eta~=~ \log 2  +  \half y   +    {1\over y}
 +  {  1\over 2 y^2}   ( 9 \log y  - 2) + O ( { \log^2 y \over y^3}), \nonumber \\
\phi~=~-{1\over 2}\log(2Q^2) + 2\log y - {9\over y}\log y + O({{\log y}\over y^2}).
\label{ktIR}
\end{eqnarray}
It the IR thus $T\rightarrow T_{IR}=0$. The Einstein-frame metric  
asymptotes to $AdS_5 \times S^5$  at $\rho=\infty$
and evolves into a metric with negative Ricci scalar at smaller $\rho$.

From the IR behaviour of $\eta$ we can deduce that:
\begin{equation}
\log\rho - \log(\log\rho)^2~\approx ~-4\log u.
\end{equation}
From this, the subsequent asymptotic behaviour for the effective kinetic term 
in the probe action, follows:
\begin{equation}
dE_k~\approx~-\half \tau_p {1\over\log u}\left(du^2 + u^2 d{\Omega_5}^2\right).
\label{IRcin}
\end{equation}
From this we see that in the IR case , the kinetic term tends to zero. This
reflects the fact that the coupling of the gauge theory tends to infinity. 
The contributions to $E_k$ coming from $K(T)$, do not modify the asymptotic behaviour 
(\ref{IRcin}), but they are crucial in determining the sign of the kinetic term.
Moving a little away from the IR extremum, it is possible to verify that the kinetic term does not become negative. This depends crucially on 
the coefficient $3\over {32}$ which multiplies the $T^2$ term 
in $K(T)$ (\ref{k(t)}).

Now let us examine the potential term for the probe. We get the following near IR
behaviour:
\begin{equation}
V~\approx~{\tau_p\over Q}\left({1\over\sqrt2} - 1\right){u^4\over(\log u)^2}.
\label{bla} 
\end{equation}
So the potential energy tends to zero from negative values. This means that  
probe calculations give a repulsive force between the sources and the probe in the IR. We can thus imagine that a distribution of branes more complicate than
the simple stack of parallel ones is the source of our bakground. The solution we are discussing here is conformal in the IR, which corresponds to fine tune the gravity solution so to have that the scalars remain massless. This should correspond to taking the D-branes forming an $SO(6)$ shell in the transverse space. We will see however that the low velocity limit do not 
seem to be a sensible one for this IR solution. So all these considerations must be
taken with caution. Let us only outline that eq. (\ref{bla}) could give indicative suggestions about the scalar potential in the IR.

The IR solution proposed for the confining case \cite{minahan} has instead the following behaviour:
\begin{eqnarray}
\phi~\approx~ \alpha_0\rho + \phi_0 ,\nonumber\\
\xi~\approx~ \alpha_1\rho + \xi_0 ,\nonumber \\
\eta~\approx~ \alpha_2\rho +\eta_0,\nonumber \\
\alpha_i>0, i=0,1,2
\label{mIR}
\end{eqnarray}
with the tachyon exponentially relaxing to zero and with:
\begin{equation}
\half{\alpha_0}^2 + \half{\alpha_1}^2 -5{\alpha_2} = 0.
\end{equation}
In \cite{minahan} was showed that the particular solution with $\alpha_1=\alpha_2$ 
and $\alpha_0 = 3\alpha_1$ could receive mild corrections from
string loops, so that the corresponding gravity solution could be used to study the
conjectured dual field theory. Wilson loop calculations showed that the solution in 
(\ref{mIR}) can give confinement from the gauge theory side. Here we want to discuss
this background from the point of view of the probe effective action.

Let us consider the solution with $\alpha_1=\alpha_2 , \alpha_0 =3\alpha_1$. The 
relation between  $\rho$ and $u$ now reads:
\begin{equation}
{\log u}\approx e^{-2\alpha_1\rho}.
\end{equation}
Just note that now, with $\rho\rightarrow\infty$, $u\rightarrow 1$. This is 
 an IR limit from the gauge theory side. In fact what we call ``energy'' 
 $E$ in the field theory context, is a parameter which, naively, can be read from the 
tension of a string stretched between the sources at $u=1$ and a D3 probe at a 
generic $u$ \cite{malda0}. Thus if we consider a string stretched along $u$ direction,
we get ($\tau = x_0 ; \sigma = u$):
\begin{equation}
E~\approx~\int du e^{\half\phi}\sqrt{g_{00}g_{uu}}~=~\int du {e^{\half (\phi - \eta)}\over u}.
\end{equation}
Just note that this is the same relation (with $\Phi$ prae E) between the new variable and $u$ which was found in writing the gauge theory lagrangian in canonical form (see eq. (\ref{PHI})).
For the solutions (\ref{mIR}) with $\alpha_1=\alpha_2 , \alpha_0 =3\alpha_1$, this gives:
\begin{equation}
E\approx \sqrt{\log u}.
\label{enrgIR}
\end{equation}
So from these naive considerations it  could be  argued that $E\rightarrow 0$ 
when $u\rightarrow 1$. The fact that the background has a singularity in $u=1$ 
\cite{minahan}, renders impossible a stronger motivation for eq. (\ref{enrgIR}), 
which we take only as an indicative relation.\footnote{It is quite simple to verify that for the UV and  the IR conformal solutions 
previously discussed, we (asymptotically) have  $u\approx E$. The same relation can
be obtained from holographic considerations, following the prescriptions of \cite{sw} and
\cite{pepo} for relating energy and radius. The same prescriptions fail however in the nonconformal case just as was noted in \cite{pepo}.} 

The kinetic term for the probe has now the following near IR behaviour:
\begin{equation}
dE_k\approx \tau_p {\sqrt{\log u}}\left(du^2 + u^2 d{\Omega_5}^2\right).
\end{equation}
We see that this term goes to zero in the IR.
For the leading terms in the potential energy we find:
\begin{equation}
V\approx \tau_p \sqrt{\log u}\approx E.
\end{equation}
The force the probe experts is thus attractive, with a modulus which tends
to infinity when $u\rightarrow 1$. This seems to imply that for the particular solution of \cite{minahan} the stack of parallel branes could be a stable
configuration, differently from the fine tuned conformal ones.

It is interesting to note that $c$, the ``speed of light in transverse space'', goes to
zero in the IR conformal case (thus requiring a little of justification in taking the
 slow velocity limit), while for the confining solutions in (\ref{mIR}) it has the following behaviour:
\begin{equation}  
c^2\approx (\log u)^{(\alpha_2-\alpha_1)\over 2\alpha_2}u^2.
\end{equation}
For the particular solution with $\alpha_1=\alpha_2$ we have considered here, we have
$c\neq 0$ in the IR, which could make the slow velocity limit more sensible.

\subsection{Probing the $N+M$ IR conformal solution}
Let us now try to extend the previous considerations to the IR conformal solution discussed at the beginning of this work. Suppose having constructed a background generated by N+M sources and moving towards them a dyonic D3 probe. The effective action for this probe reads:
\begin{equation}
S_p~=~ -2\tau_p \int d^4\xi e^{-\phi}\sqrt{-\hat{G}} -2g^{-1}_s\mu_p\int d^4\xi\left[ Q \int d\rho e^{-2\xi}f^{-1}(T) + P \int d\rho e^{-2\xi}f(T)\right].
\label{dyprobe}
\end{equation}
We have used here the fact that the terms proportional to T in $K_{el}(T)$ and $K_{mag}(T)$, cancel each other (see \cite{kt1}). Moreover, as argued in \cite{kt1}, we have set $f^{-1}_{el}(T) = f_{mag}(T)$.

From (\ref{dyprobe}) we immediately see that the kinetic term for this probe is
simply two times the one in (\ref{IRcin}). For the potential energy we get, 
using (\ref{eqgen}):
\begin{equation}
V~\approx~ {{\sqrt {2}\tau_p}\over{\sqrt {(Q^2 + P^2)}}}\left[ 1- {(Q+ P)\over{\sqrt{2(Q^2 + P^2)}}}\right]{u^4\over(\log u)^2} .
\end{equation}
In the dyonic case $Q=P$ of \cite{kt1}, in which the background is exactly $AdS_5\times S^5, T=0$, the effective mass of a dyonic probe is constant and the potential is zero. For $Q\neq P$, $V$ is always positive, unlike in the purely electric case. This implies that the dyonic probe experts an attractive 
force. 

If we consider just an electric  probe we get, for the potential:
\begin{equation}
V~\approx~ {\tau_p\over \sqrt {2(Q^2 + P^2)}}\left[1- {2Q\over{\sqrt{2(Q^2 + P^2)}}}\right]{u^4\over(\log u)^2}.
\end{equation}
So we can have $V>0$ only if $P>Q$. For a magnetic probe the same is true if $Q>P$. Thus the stack of $N+M$ branes is unstable with respect to moving off
an electric or magnetic brane.

There exist thus instabilities in the fine tuned background we have constructed. This suggests that a different assumption, instead of the naive one in which the branes are taken parallel to each other, should be taken.
\section{Discussion}
It would be interesting to extend this kind of analysis to orbifolds and orientifolds
of Type 0 string theory. For example a modified $\Omega$ projection exists,
which removes the closed string tachyon (and modifies the closed string spectrum which becomes chiral), allowing at the same time
an open $U(32)$ spectrum free of tachyons as well \cite{sagnotti}.
\footnote{We would like to thank A. Sagnotti for having reminded us this fact}. See also \cite{dudas} for brane constructions in this context.

It would also be interesting to make explicit the angular momentum of the probe in our 
calculations. As the probe moves on geodesics, this is a conserved quantity. Effects
of the angular momentum in cosmological analysis of Type 0 backgrounds were studied in
\cite{pap}. For a study of possible singularity resolutions via angular momentum in the
context of Type II theory, see \cite{malda}.

\vskip .2in \noindent \textbf{Acknowledgments}\vskip .1in \noindent  
A special thank to A. Zaffaroni for suggesting part of the investigations here
reported and for  constant help and  suggestions. Thanks to A. Bigazzi, 
L. Girardello, A. Pasquinucci and A. Refolli for useful discussions and suggestions.  
F. B. is partially supported by INFN and MURST, and
by the European Commission TMR program HPRN-CT-2000-00131, 
wherein he is associated to the University of Padova.

\end{document}